\documentclass[apj, twocolumn, numberedappendix, appendixfloats, twocolappendix]{openjournal}

\newcommand{\e}[1]{\times 10^{#1}}
\newcommand{\msun}{\rm{M_\odot}}

\usepackage{graphicx}	
\usepackage[dvipsnames]{xcolor}

\usepackage{amsmath}	
\usepackage{multirow}
\usepackage{enumitem}
\usepackage[breaklinks,colorlinks,citecolor=blue,urlcolor=blue]{hyperref}
\usepackage{orcidlink}

\shorttitle{Light Cone Strong Lensing}
\shortauthors{Roche et al.}

\begin{document}

\title{\vspace{-0.5cm}A Consistent Implementation of Cluster Strong Lensing in Cosmological Simulation Light Cones\vspace{-1.5cm}}


\author{Cian Roche\,\orcidlink{0000-0002-3400-6991}$^{1,*}$}
\author{Mark Vogelsberger\,\orcidlink{0000-0001-8593-7692}$^{1,2}$}
\author{Michael McDonald\,\orcidlink{0000-0001-5226-8349}$^{1}$}
\author{Isaque Dutra\,\orcidlink{0000-0001-7040-4930}$^{3}$}
\author{Priyamvada Natarajan\,\orcidlink{0000-0002-5554-8896}$^{3,4}$}
\author{Xuejian Shen\,\orcidlink{0000-0002-6196-823X}$^{1}$}
\author{Soumya Shreeram\,\orcidlink{0000-0001-9789-5362}$^{5}$}
\author{R. Benton Metcalf\,\orcidlink{0000-0003-3167-2574}$^{6,7}$}
\author{Keren Sharon\,\orcidlink{0000-0002-7559-0864}$^{8}$}
\author{Simon Birrer\,\orcidlink{0000-0003-3195-5507}$^{9}$}
\author{Wonki Lee\,\orcidlink{0000-0002-1566-5094}$^{10}$}
\author{Massimo Meneghetti\,\orcidlink{0000-0003-1225-7084}$^{7}$}

\affiliation{$^{1}$Department of Physics and MIT Kavli Institute for Astrophysics and Space Research, \\
77 Massachusetts Avenue, Cambridge, MA 02139, USA}
\affiliation{$^{2}$The NSF AI Institute for Artificial Intelligence and Fundamental Interactions, \\
77 Massachusetts Avenue, Cambridge, MA 02139, USA}
\affiliation{$^{3}$Department of Physics, Yale University, New Haven, CT 06511, USA}
\affiliation{$^{4}$Department of Astronomy, Yale University, New Haven, CT 06511, USA}
\affiliation{$^{5}$Max Planck Institute for Extraterrestrial Physics, Gie\ss enbachstraße 1, 85748 Garching bei München, Germany}
\affiliation{$^{6}$Dipartimento di Fisica e Astronomia ``Augusto Righi" - Alma
Mater Studiorum Università di Bologna, via Piero Gobetti 93/2,
40129 Bologna, Italy
}
\affiliation{$^{7}$INAF-Osservatorio di Astrofisica e Scienza dello Spazio di
Bologna, Via Piero Gobetti 93/3, 40129 Bologna, Italy}
\affiliation{$^{8}$Department of Astronomy, University of Michigan, 1085 South University Avenue, Ann Arbor, MI 48109, USA}
\affiliation{$^{9}$Department of Physics and Astronomy, Stony Brook University, Stony Brook, NY 11794, USA}
\affiliation{$^{10}$ Yonsei University, Department of Astronomy, Seoul, Republic of Korea}

\thanks{$^*$E-mail: \href{mailto:roche@mit.edu}{roche@mit.edu}}

\begin{abstract}
Galaxy cluster strong gravitational lensing plays a central role in precision cosmology, yet robust theoretical predictions have lagged behind an abundance of high-quality strong lensing observations. This shortfall reflects both a mismatch between the geometry of the strong-lensing problem and standard cubic simulation boxes, and the fundamental tension between simulation volume and resolution. Consequently, many current forecasts adopt hybrid approaches that extract individual lenses from simulations and combine them with analytic or observed source populations positioned near caustics. These methods often omit correlated and/or uncorrelated line of sight (LoS) structure, or include it in ways that do not preserve correlations across redshift. Here we present a fully simulation-based procedure that generates strong-lensing images directly from particle data, drawing the lens, source, and all intervening resolved objects self-consistently from the simulated large-scale structure. Our approach combines a structure-preserving remapping of the simulation volume into a lensing-appropriate geometry with multi-plane ray tracing, enabling the use of uniform simulation boxes that resolve both cluster-scale primary lenses and high-redshift source galaxies. We demonstrate the method by generating example light cones and images using IllustrisTNG data, then use these results to conservatively quantify the impact of LoS structure on image configurations and critical-curve morphology. We find that uncorrelated LoS structure can shift the relative positions of lensed images by several arcseconds, introduces a $\sim 6\%$ scatter in the area of a cluster’s primary critical curve, and changes the total critical area within 100$\arcsec$ of the cluster potential minimum by $16^{+20\%}_{-14\%}$ at a source plane redshift of $z_s=4$. These results highlight the importance of incorporating full light-cone information in theoretical predictions of strong lensing, and of carefully accounting for line of sight structure when interpreting observations. 
\end{abstract}

\keywords{\href{http://astrothesaurus.org/uat/1643}{Strong gravitational lensing(1643)}; \href{http://astrothesaurus.org/uat/265}{Cold dark matter(265)}; \href{http://astrothesaurus.org/uat/584}{Galaxy clusters(584)}; \href{http://astrothesaurus.org/uat/767}{Hydrodynamical simulations(767)}}

\section{Introduction}\label{sec:intro}
Gravitational lensing has proven to be a powerful tool to probe the matter distribution in the universe. On the largest scales, lensing of the CMB informs cosmological parameters and the large-scale structure of the universe \citep{lewis:2006:cmblensing,planck2020:cmb_lensing}. On galaxy cluster scales, weak and strong lensing permit mapping of the predominantly dark matter mass distribution independent of any assumptions about the dynamical state of the lensing cluster \citep{umetsu2020:weak_lensing, niemiec2023:weakstrong, Natarajan:2024, cerny2025:mass_profiles}. Cluster lenses serve as cosmic telescopes and bring into view the first stars and other faint background objects that would otherwise be undetectable \citep{bradley2023:cosmic_telescope}, and enable detailed studies of the smallest building blocks of magnified galaxies at cosmic noon \citep{florian2021}. On galaxy scales, meanwhile, strongly lensed background galaxies and quasars have enabled measurements of the Hubble constant \citep{birrer2019:h0licow, tdcosmo2025}, and perturbations to lensed arcs have been used to detect small-scale substructure \citep{vegetti2012:substructure, hezaveh2016:substructure}, some of which appears to be extremely dense, suggestive of a potential signature of dark matter self-interaction physics \citep{nadler2023:sidm_substructure, despali2022:substructure, despali2025:substructure2, powell2025:substructure, minor2025:substructure, vegetti2026:substructure}. On the smallest scales, gravitational microlensing has been used to rule out large portions of parameter space for Massive Astrophysical Compact Halo Objects (MACHOs) as a putative dark matter candidate \citep{alcock2000:MACHOs, tisserand2007:MACHOs, wyrzykowski2010:MACHOs}, and to discover many exoplanets in a region of parameter space largely inaccessible to other detection methods \citep{penny2019:microlensing}. Additionally, strong lensing of gravitational waves poses significant promise in measuring the Hubble constant and localizing gravitational wave sources \citep{shan:2025:gwlensing}. \\

\begin{figure*}
    \centering
    \includegraphics[width=\linewidth]{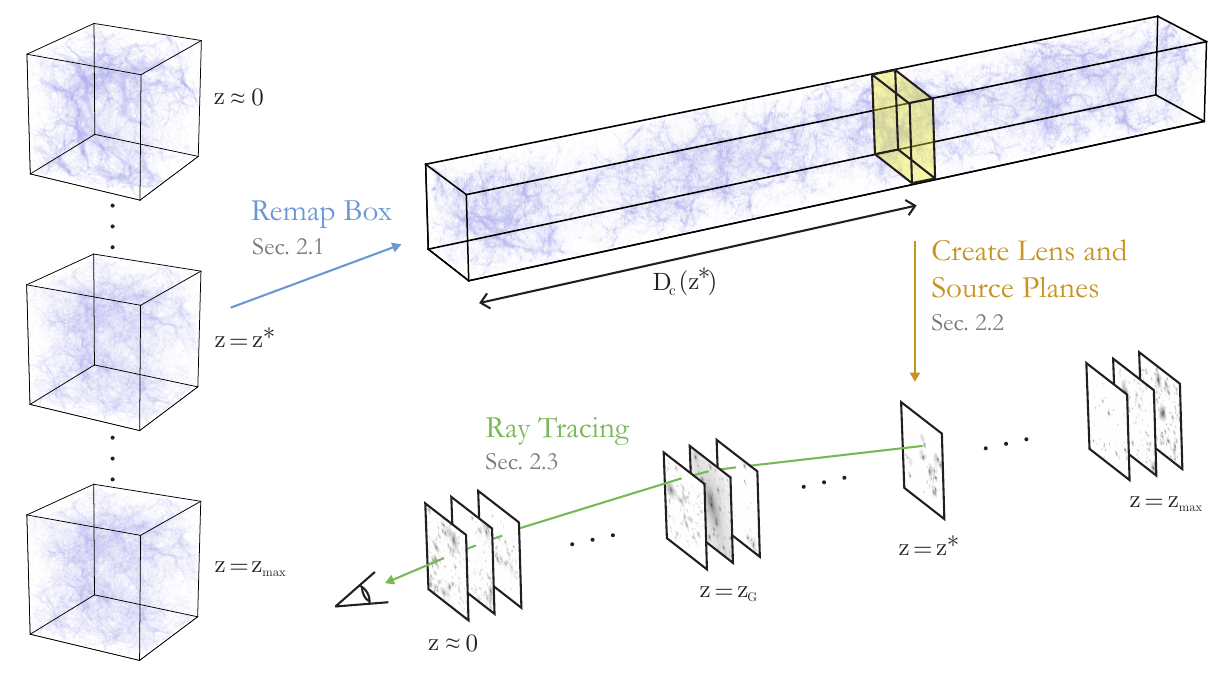}
    \caption{Schematic of the procedure by which many snapshots of a cosmological simulation run in a cubic box can be combined into a multi-plane strong lensing image. One source and one lens plane are created per simulation snapshot (i.e. redshift), and then the light from each source plane at redshift $z_s$ is ray traced through all lens planes with redshift $z_l$ such that $z_l < z_s$. The example plane shown in the diagram is at a higher redshift than the primary lens group at $z_G$, but the planes at $z<z_G$ are also ray traced in this pipeline. The resulting lensed image is then the sum of all lensed source planes. Observational effects can then be modeled and applied to the result for various instrument specifications. The simulation data shown in this example is a portion of the TNG50 box.}
    \label{fig:toc}
\end{figure*}

The strong gravitational lensing regime, characterized by the presence of multiple images, arcs, and/or rings, has been particularly valuable in investigating the mass distribution of galaxies and galaxy clusters and is entering an era of both unprecedented high resolution through JWST, and unprecedented statistical power through Euclid, the ongoing Legacy Survey of Space and Time (LSST), and the Nancy Grace Roman Space Telescope. However, the interpretation of strong gravitational lensing observations is highly complex, requiring the identification of multiple image configurations, measuring their redshifts, employing an appropriate method for representing the mass distributions analytically, and developing the theoretical predictions necessary to understand the dark matter and baryonic physics that undergird the observations. \\

Historically, such predictions have been created using analytical mass profiles placed in one or more lensing planes, behind which an analytical point source is positioned. The rays from this source are then traced to an observer to create a lensed image, and the relevant lensing quantities can be studied as a function of parameters like the geometrical configuration, or the choice of analytical profiles, which can be motivated by differing dark matter or baryonic physics models \citep{bartelmann2010:lensing_review}. As the quality and size of the observational ensemble of strong lensing systems has grown, these predictions have become significantly more sophisticated. Advancements have included distributing the mass profiles of the primary lens according to the statistics of subhalo catalogs of cosmological simulations \citep{vogelsberger:2020:review}, including line of sight substructure drawn from the same catalogs, replacing the analytical profiles for the primary lens with pixelated mass distributions or particle data extracted from simulations \citep{meneghetti2017:challenge}, and replacing the analytical sources and line of sight objects with real observations of galaxies \citep{li:2016:pics}, for example from the HST COSMOS sample \citep{koekmoer2007:cosmos}. These improvements have enabled the study of more complex lensing questions, such as giant arc perturbations and substructure as a probe of dark matter physics \citep{powell2025:substructure}, and even searching for dark halos without a luminous counterpart \citep{wagnercarena:2024:substructure}. Notably, however, certain foundational questions about these predictions remain debated in the literature, such as the known deficit of galaxy-scale lensing in isolated galaxies \citep{li2002:galaxy_lensing} and within clusters \citep{meneghetti2020, meneghetti2022, meneghetti2023, tokayer2024, dutra2025:ggsl} relative to predictions from cosmological simulations, and the exact role of the arcsecond-scale relative deflections \citep{host2012:los_lensing, raney2020:los_lensing} and arcminute-scale absolute deflections \citep{uros1994:LSS_lensing} due to line of sight material and large scale structure when interpreting observations and generating predictions. \\

Despite the extensive literature surrounding simulated strong lensing and cosmological simulations, there do not yet exist simulated strong lens images produced consistently from the outputs of simulations. By ``consistently", we mean that the primary lens, sources, and all intervening matter along the line of sight are drawn from the same simulation, in a manner faithful to the clustered structures of that simulation as predicted to form and evolve per the standard cold dark matter paradigm of structure formation. Light cones have been produced in cosmological simulations in various geometries, most completely in MillenniumTNG \citep{hernandez:2023:mtng, pakmor2023:mtng, barrera:2023:mtng}, but limitations of the simulation box size necessitate tiling the box several times to reach the desired source redshifts for such light cone products. This approach has some drawbacks, such as introducing the possibility of the same structure being present multiple times in the same field of view if the line of sight vector is not chosen carefully \citep{zhao2024:box_repetition}, and the matter power spectrum of the light cone being biased on scales similar to and larger than the initial box size. This box repetition technique has been successfully used in the literature both for weak lensing \citep{gouin2019:WL_light_cone, hernandez:2023:mtng} and for strong lensing, for example by \cite{hilbert2007:strong_lensing_optical_depths}, who ray trace through light cones in the N-body Millennium simulation \citep{springel:2005:millennium} to study strong lensing optical depths in $\Lambda{\rm CDM}$. Mock light images of strong lensing clusters have not been produced directly from simulation light cones however, due to the tension in computational resources between the volume necessary to produce large clusters, and the resolution required to resolve the stellar content of background galaxies and galaxy central concentrations to which strong lensing is sensitive \citep{treu2010:sl_review, fox2022:core_shape}. \\

In this paper, we demonstrate a method for adapting simulation data to the geometry of strong lensing, without utilizing repeated structures, by first remapping the whole box into a shape suitable for the task. This remapping is computed for each snapshot of the simulation, and planes of mass and light are extracted at the corresponding redshift, enabling the construction of a light cone in which multi-plane ray tracing can be performed. This approach requires almost no manual intervention, sampling, or decisions beyond the assumptions of the simulation cosmology and feedback prescription, and avoids the issue of repeated structure without careful choice of the line of sight. This simulation volume remapping technique was first introduced in \cite{carlson2010:boxremap}, and has since been utilized in various fields, for example to produce mock surveys comparable to observations studying clustering \citep{manera2013:boss_mocks, torre2013:VIPERS_mocks}, baryon acoustic oscillation measurements \citep{kazin2014:bao, metin2018:bao}, weak lensing \citep{giocoli2016:weak_lensing_mocks, zhao2024:box_repetition}, neutrino masses \citep{thiele2024:neutrino_masses} and studies of the circumgalactic medium \citep{shreeram2025:methods, shreeram2025:LSS_CGM}. We apply this methodology for the first time to the cluster strong lensing problem, and make use of current-generation cosmological hydrodynamical simulations whose box size and resolution enable the generation of consistent strong lensing images when combined with the box remapping approach. \\

In Sec.~\ref{sec:remap} we discuss the challenges of the lensing problem for cosmological simulations and describe how the technique of volume remapping addresses these issues. In Sec.~\ref{sec:planes} we outline the methods we use to turn simulation particle data into source (light) and lens (mass) planes, and describe the ray tracing procedure used to produce the final images and maps of lensing quantities in Sec.~\ref{sec:raytracing}. In Sec.~\ref{sec:data} we provide the simulation details for IllustrisTNG, which we use for the demonstration of the method, and show some fiducial results of this procedure in Sec.~\ref{sec:results}, studying in particular the critical curve structure and impact of the line of sight material. We conclude in Sec.~\ref{sec:conclusions} with an outlook for the use cases for this methodology for cluster studies as well as time delay cosmography studies, and reiterate the limitations of the methodology.\\

\section{Methodology}\label{sec:methodology}
\begin{figure*}
    \centering
    \includegraphics[width=\linewidth]{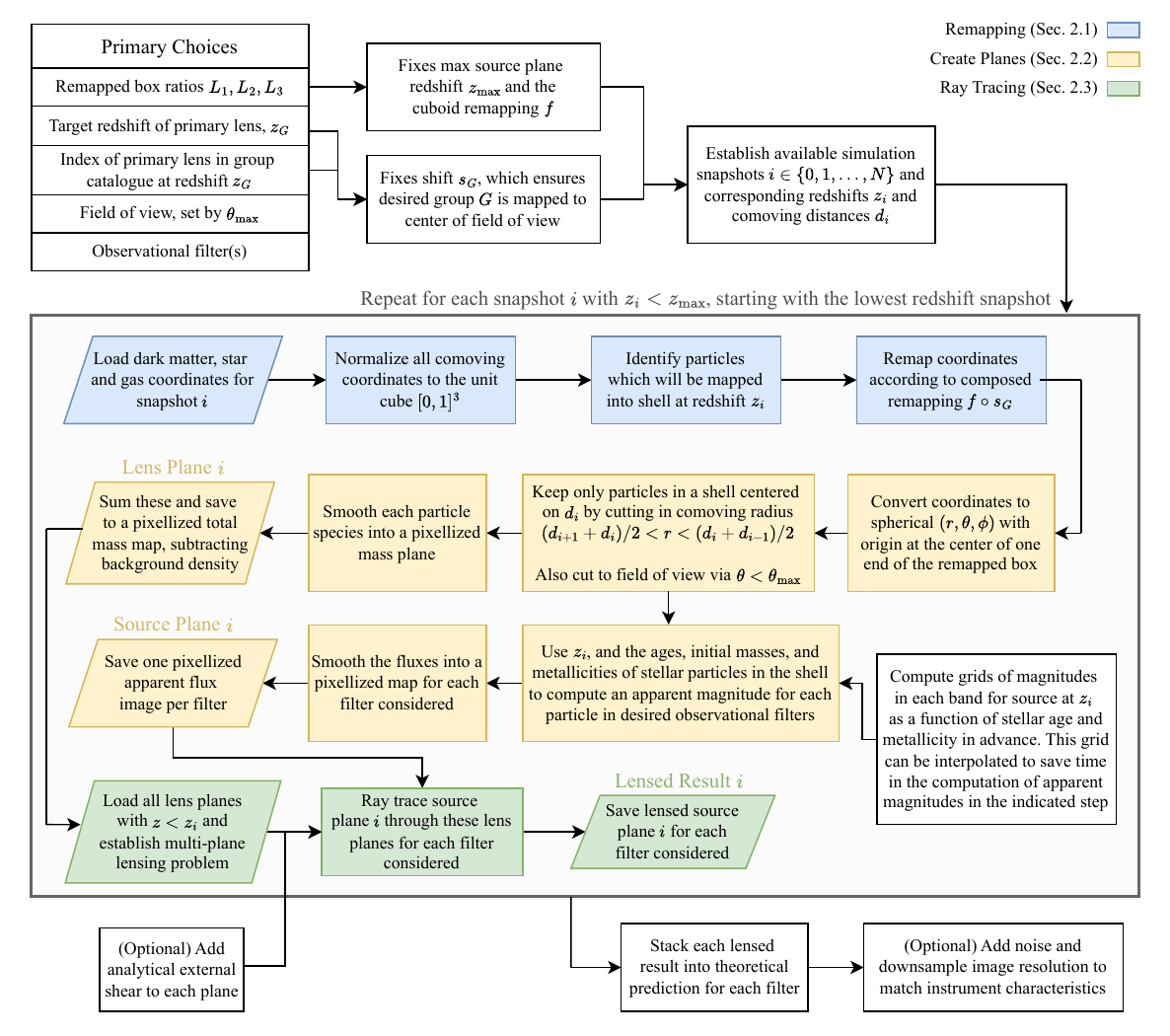}
    \caption{A flowchart of the computation required for a single lensed image prediction using a group $G$ from a cosmological simulation as the primary lens at a desired redshift $z_G$. Note that lens plane $i$ is not used in the computation of lensed result $i$ and this arrow is included only to direct visuals.}
    \label{fig:flowchart}
\end{figure*}

In this section, we describe the procedure by which we generate light cones from cosmological simulations and ray trace through these cones to produce strong lensing images. A high-level overview of this process is shown in Fig.~\ref{fig:toc}, followed by a more detailed flowchart in Fig.~\ref{fig:flowchart}. This flowchart highlights in particular that the number of assumptions necessary to produce a strong lensing image is less than typically used; for example it does not include a choice of halo mass function, analytical mass profiles, or types of sources injected in caustics. Instead, a cosmological hydrodynamical simulation volume, on which choices of mass function and profile shape are often based, is used as the prior on cosmology and galaxy formation model directly. 

\subsection{Box Remapping} \label{sec:remap}

One of the primary challenges in generating light cone strong lensing images from cosmological simulation data is the distance to typical cluster-scale lensed sources, such as background galaxies or quasars \citep{Natarajan:2024}. Most source objects for cluster lenses are found between redshift $\sim$1--4 \citep{bayliss2011:sl_source_redshifts} arising from a mixture of geometric favorability in the lensing configuration and the redshift dependence of source number density\footnote{Though a significant fraction of sources can be at redshifts $>4$ for more distant cluster lenses \citep{boldrin2012:euclid_giant_arc_stats}.}, with $z=4$ corresponding to a comoving distance of $\sim7.5\,{\rm cGpc}$. To generate a single lensed image with lensed background sources, it is therefore necessary to use a simulation with a box side length close to $\sim7.5\,{\rm cGpc}$. Simultaneously, since strong lensing is known to depend on the inner structure of the mass profile of the lens objects \citep{fox2022:core_shape}, high-resolution simulations capable of resolving small-scale sources and the cluster cores responsible for the lensing must be used. These two conditions together necessitate a simulation that does not yet exist, and is not feasible for the near future due to tremendous computational requirements. For example, the state-of-the-art MillenniumTNG simulation utilizes a 500~cMpc$\,h^{-1}$ box at a mass resolution that can confidently resolve\footnote{Objects containing at least 200 dark matter particles.} down to a mass of $\sim 4\times10^{10}\,{\rm M_\odot}$, and spatial features on $\sim5\,{\rm kpc}$ scales. These simulations required $1.7\times 10^8$ core-hours to run, and over $1\,{\rm PB}$ of storage \citep{pakmor2023:mtng}. To maintain this same mass resolution (which already cannot resolve smaller line of sight material and background sources) but increase the box size to $\sim7.5\,{\rm cGpc}$ would require $> 1000$ times the computing resources, which no existing or planned computing facility possesses.\\

In contrast to the geometrical challenge of the line of sight length required for consistent lensing, an opportunity is present in the field of view. Strong lensing observations occupy fields of view no larger than a few hundred arc seconds, or in the rest frame of a distant source at redshift $z=4$, a length on the order of only a few Mpc. If instead of a cosmological simulation run in a cubic box, we utilized a box geometry of approximately a right square prism (square cross section and one long axis, see Fig.~\ref{fig:toc}) this geometry would be ideal to generate consistent strong lensing images. However, practical uses for such a simulation would be limited, and some numerical challenges exist when choosing extreme axis ratios as would be favorable for the strong lensing problem \citep{carlson2010:boxremap}. As such this simulation data does not yet exist. \\

It is possible, however, to replicate this geometry for existing simulations using the technique of a \textit{box remapping}. This technique relies upon a mapping from a cubic simulation box with periodic boundaries to a box with different axis ratios and periodicity along only one axis. For this purpose, we use the code \texttt{BoxRemap} \citep{carlson2010:boxremap}, particularly the C++ implementation, as we will be remapping a large number of particle positions for each produced image. This remapping preserves the total volume, local structure, uses the information for each particle in the original simulation exactly once, and does not map distant structures in the initial box to be close together in the remapped box. Formally, this is achieved by creating a lattice tiled in 3 dimensions with the initial simulation box and finding a unit cell for this lattice in the shape of a rectangular cuboid with the desired axis ratios between the initial and remapped boxes. We label these axis ratios $L_1$, $L_2$ and $L_3$, and note that only certain solutions for the $L_j$ are valid, meaning one must first find these solutions and then choose the solution most closely matching the desired ratios. One can then develop\footnote{For the details of this transformation, see \cite{carlson2010:boxremap}. Note that a package for lightcone creation with this transformation using TNG data has been implemented in \href{https://github.com/SoumyaShreeram/LightGen}{LightGen}, but this package is not used in this paper.} an invertible transformation $f : [0,1]^3 \rightarrow [0,L_1]\times[0,L_2]\times[0,L_3]$ to map points from the unit cube to a cuboid with the desired properties. For application to simulation data, first the coordinates can be normalized by the box size, then after remapping this normalization can be undone. \\

We choose the dimensions of the remapped box according to our needs for lensing and to address the limitations of the initial box size. We seek remapped axis ratios with two short and approximately equal axes ($L_1$ and $L_2$, the plane of the sky) and one long axis ($L_3$, for the line of sight). For a desired maximum source plane redshift of $z_{\rm max}\simeq 4$, i.e. a distance of $\sim7.5{\rm\,cGpc}$, and an initial simulation box side length of roughly $300{\rm\,Mpc}$ (as it is for TNG300-1), we require an $L_3$ of approximately $25$. This number cannot be arbitrarily increased, however, due to our requirements for the field of view. For example, for a desired square field of view of $200''$, and a maximum source plane redshift of $z_{\rm max}=4$, we require that the smallest remapped box side length be $\gtrsim 2{\rm\,cMpc}$. For an initial box side length of $300{\rm\,cMpc}$ this corresponds to a minimum axis ratio of approximately $L_1,L_2 > 0.005$. In practice, some spurious correlations can appear at the boundary regions of the remapped box \citep{carlson2010:boxremap}, and so choosing a remapped geometry with a larger field of view than necessary is advantageous, so these regions will not appear in the generated light cones. Finally, since the mapping is volume-preserving, we require $L_1L_2L_3 = 1$. If we fix the line of sight length ratio as $L_3=25$, then the ideal ratios $L_1$ and $L_2$ for a square field of view would satisfy $L_1 \simeq L_2 = \sqrt{1/25} = 0.2$.  We use the \texttt{genremap} utility within \texttt{BoxRemap} to find solutions for $L_{1,2,3}$ that most closely match these conditions, and find a solution with our fiducial axis ratios of $L_1=0.2$, $L_2=0.2$, and $L_3=25$. These ratios will be used for all images produced in Sec. \ref{sec:results}, but can be adjusted based on the needs of a particular analysis. These axis ratios can be represented by the following transformation matrix, 
\begin{equation*}\label{eq:matrix}
    \begin{pmatrix}
    u_{11} & u_{12} & u_{13} \\
    u_{21} & u_{22} & u_{23} \\
    u_{31} & u_{32} & u_{33} 
    \end{pmatrix}
    =
    \begin{pmatrix}
    4 & 4 & 3 \\
    1 & 0 & 1 \\
    16 & 15 & 12 
    \end{pmatrix}
\end{equation*}

in the notation of \cite{carlson2010:boxremap}, noting that \texttt{genremap} assumes the $\hat{e}_1$ direction to be the ``long" axis and so an additional transformation is required to permute the basis vectors. With $L_3=25$ and a box size of $205\,{\rm Mpc}\,h^{-1}$, we find a $z_{\rm max}$ of $\sim4.24$, such that the highest redshift snapshot available to be considered in the data described in Sec.~\ref{sec:data} is $z_{\rm max}=4.18$. The light cones in this paper will therefore be created using source and lens planes at redshifts no larger than $4.18$. Note that for a box size of 500~cMpc$\,h^{-1}$ as for MillenniumTNG, these remapped box ratios would enable a light cone with arbitrarily high $z_{\rm max}$, with the added advantage that the larger volume would sample a larger variety of structures across a collection of many light cones. This advantage is relevant depending on the needs of a particular study, but for this demonstration of method here, we use TNG300 due to the smaller data volume. \\

Another consideration for generating lensing images is that we wish for massive clusters to be present in the center of the field of view and to be at redshifts relevant for strong lensing to occur efficiently, to adequately mimic targeted observations of cluster lens fields. Applying the transformation $f$ to the simulation box would result in massive clusters distributed throughout the remapped volume, perhaps close to the edges and not at appropriate redshifts for strong lensing observations by some chosen observer. To avoid this, we first choose the desired location of a massive cluster (group)\footnote{We will frequently refer to galaxy clusters as ``groups" in the remainder of the text, reflecting the terminology used to identify structures in simulations \citep{springel:2005:gadget}.} of interest in the remapped box. For simplicity, we choose the target location in the remapped box to be ($L_1/2, L_2/2$) in normalized coordinates in the plane of the sky ($\hat{e}_1$ and $\hat{e}_2$ directions) and at the normalized coordinate in the $\hat{e}_3$ direction corresponding to some target cluster redshift $z_G$, labeled $d_{z_G}$. This redshift fixed at $z_G=0.3802$ for each image produced in this paper, chosen to match \cite{robertson2020:einstein_radii} which we will later compare to, with the specific value being set by the closest simulation snapshot redshift to 0.38. Note however that this value can be adjusted or sampled from a distribution depending on the specific use case.\\

We use the inverse of the remapping $f^{-1}$ to find the corresponding location $\vec{x_0} = f^{-1}((L_1/2,L_2/2,d_{z_G}))$ in the initial box. This coordinate in the original box will be remapped to the desired position, and so we exploit the periodicity of the initial box to shift the coordinates of all particles in order to place a chosen group $G$ at the position $\vec{x_0}$. This is accomplished by finding the position $\vec{x}_G$ of the group $G$ in the simulation snapshot at redshift $z_G$, and shifting all coordinates by the mapping $s_G:\vec{x} \rightarrow \vec{x} + \vec{x}_0 - \vec{x}_G$. We then use the combined transformation $f \circ s_G$ to remap the box (working in comoving coordinates) at all redshifts/snapshots consistently for a single target group, which allows us to produce the lens and source planes described in Sec. \ref{sec:planes} that will form the light cone.\\

We perform this procedure for every group $G$ of interest. This results in a small chance for some objects to be present in multiple light cones (each with different target groups), but the same structure cannot be present multiple times in a single light cone. The fraction of the volume of the simulation box present in a single light cone is very small ($\sim 0.01\%$ for a $200\arcsec$ field of view up to $z_{\rm max}=4$ in TNG300-1), so the probability that an object appears in multiple light cones is small even for moderately-sized catalogs of light cones. A very large catalog of many images produced using this method could thus exhibit some non-physical correlations between images, but the relevance of these correlations to lens modeling and population studies is outside the scope of this current work. 

\subsection{Creating the lens and source planes} \label{sec:planes} 
Once the remapping for a particular group G is established, where G is the index of a chosen primary lens cluster in the group catalog of the simulation at the chosen redshift $z_G$, the simulation box can be remapped consistently at every snapshot. We now describe the procedure by which we extract mass and light information from these remapped boxes to produce a lightcone which will be used later in the ray tracing calculations (Sec.~\ref{sec:raytracing}). We utilize the standard approximation of \textit{multi-plane lensing}, which computes lensing deflections at a set of $N$ lensing planes only, rather than evaluating the cumulative effect of deflections continuously along the ray path from the full three-dimensional particle distribution (as implemented using a modified tree algorithm in \texttt{GLAMER} for example). This approach has been shown to be well converged with an inter-plane spacing below $\sim 300\,{\rm cMpc}$ \citep{petkova2014:glamer2}. In the present analysis, we use a variable interplane spacing that is always smaller than this value, corresponding exactly to the snapshot output frequency of the simulation. We discuss the snapshot frequency in more detail in Sec.~\ref{sec:data}. \\

\begin{figure*}
    \centering
    \includegraphics[width=\linewidth]{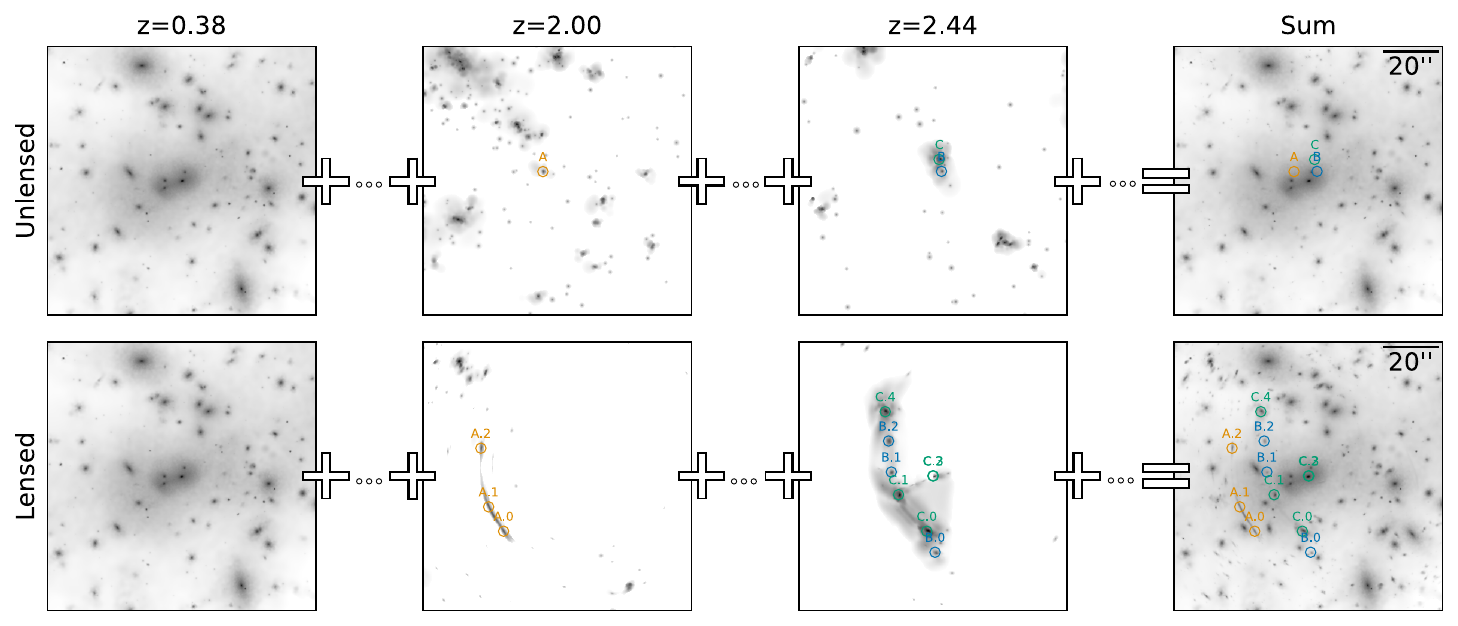}
    \caption{An illustrative example of the outputs of the lensing pipeline. \emph{(Top Row)} The surface brightness maps of 3 example planes in the light cone without lensing that surface brightness through the mass planes at lower redshifts. \emph{(Bottom Row)} How each plane looks after ray tracing the maps through all lens planes at lower redshifts. Also marked is the position of any source in the unlensed panels which is multiply imaged in the lensed panels, with each image location shown in the same color and with an alphanumeric label. Some of these multiple images are very close together and may not be visually distinguishable. On the right are the sum of surface brightnesses for all planes both unlensed and lensed, with the same multiply imaged sources marked, and the brightness of the two example planes at $z=2.00$ and $z=2.44$ enhanced by a factor of 100 for visualization in this example. Note that there exist other multiple images in the summed lensed panel which are not marked here. The lowest redshift shown is the cluster redshift, but lower redshift planes were included in the ray tracing and the sum in both rows.}
    \label{fig:individual_planes}
\end{figure*}

Recall that this observer is placed at the end of the box in the center of the plane ``looking'' along the long axis at the cluster of interest (Fig.~\ref{fig:toc}). It is therefore useful to shift the origin from $(0,0,0)$ in the remapped box to $(L_1/2, L_2/2, 0)$ in normalized comoving coordinates. In comoving spherical coordinates ($r,\theta, \phi$), a light cone is simply a cone, being defined by $r\in[0,r_{max}]$, $\theta \in [0, \theta_{max}]$ and $\phi \in [0, 2\pi]$ where $r_{max}$ is set by the maximum redshift we choose in the remapping stage (Sec.~\ref{sec:remap}), and $\theta_{max}$ is half the choice of angular field of view. Note that cutting the particle data according to this angular condition can be preferable to extracting a square field of view, as depending on the map area and cluster density profile, this choice can introduce a significant artifact in the resulting shear field known as a ``boundary truncation effect" \citep{Vyvere2020:truncation, he2023:subhalo_detection}. A second boundary truncation effect comes from using a field of view for the mass maps and ray tracing insufficiently large compared to $\theta_E$ for the primary lens. It has been demonstrated in \cite{Vyvere2020:truncation} that for galaxy-scale lenses, the artificial shear induced by boundary truncation is reduced to $\gamma_{\rm art} < 0.001$ for a field of view of $50\theta_E$. The magnitude of this effect has not been established on cluster scales where $\theta_E\sim 10$s of arcsec, but we see no significant change in results for fields of view of $500\arcsec$ and larger.  \\ 

For a desired square field of view in the produced lens and source planes of $X\arcsec$ (and thus half-width of $X/2$), this necessitates a $\theta_{\rm max}$ of at least $\sqrt{2} (X/2)$ such that the desired square field is everywhere inside the circular cut, with each corner of the square exactly meeting the circle. It is also necessary to include a buffer region in the conical cut on the particles so that particles outside the field of view can still contribute to the smoothing calculation, especially near the box corners. We choose to implement this cut as a $25\%$ increase in $\theta_{\rm max}$ such that $\theta_{\rm max} = 1.25X/\sqrt{2}$, though many choices are valid here. For the figures produced in the remainder of this paper, we fix $X=500\arcsec$ for the lens planes, and $X=100\arcsec$ for the source planes, as this is the field of view with which we will typically visualize the results. Note however that this choice restricts the largest scale mode of the matter density field which can be present in the light cone in directions perpendicular to the line of sight. Furthermore, there is a limit beyond computational cost in choosing the size of the field of view, set by the chosen remapped box axis ratios (Sec.~\ref{sec:remap}). \\

Let the list of redshifts for the available snapshots of the simulation be $\{z_i \mid i = 0, 1, \dots, N\}$ where more snapshots may exist with $z>z_N$, but no snapshots need be considered beyond $z_{\rm max}$ ($z_N \leq z_{\rm max}$, see Sec.~\ref{sec:remap}). Let the comoving distance corresponding to redshift $z_i$ be $d_i$. For each snapshot, we extract the particles in this cone and divide them into segments or ``shells'' in the radial coordinate, with the widths of the segments depending on the separation in time of the snapshot outputs of the simulation (see Fig.~\ref{fig:toc}). For a snapshot at redshift $z_i$, we take the position, mass and light information from all particles in the remapped box with radial coordinate $(d_{i-1} + d_i)/2 < r < (d_i + d_{i+1})/2$, noting that this may be asymmetric about $d_i$. This information is then used to generate a lens plane and several source planes, one for each observational filter considered. The lens plane is a combination of the mass of all dark matter, star and gas particles in the shell, while each source plane is computed using information about the ages and metallicities of the stars.\\

In practice, for each simulation snapshot $i$ considered we identify the particles in the shell centered on $d_i$ by applying $f \circ s_G$ to all particles in the snapshot, converting to spherical coordinates, and applying the cuts in $r$ and $\theta$ described above (see Fig.~\ref{fig:flowchart}). To avoid the scenario where a group lies only partially inside the shell, we also remap all group positions in the snapshot and include all particles from groups whose center lies within the shell. One remaining issue is that a group very close to the boundary between two redshift shells can, if its velocity is directed toward the shell boundary, appear again at a very similar projected position in another snapshot's shell. This is a result of the discrete nature of the simulation snapshot outputs, and would lead to the object appearing duplicated in the mass and light maps. This is expected to be rare, but could be mitigated by implementing a check for the expected position of each group near a boundary by the time of the next snapshot, and excluding it if it will likely cross the boundary.\\

Many of the remapped particles will be in sparse regions, not a part of any converged object (taking a converged object to have a total mass of at least 200 times the dark matter particle mass), which we will dub ``diffuse mass". The diffuse mass fraction increases with redshift, with up to $\sim90\%$ of the mass in shells at redshifts close to $z_{\rm max}$ being diffuse. We create the mass maps using only the particles in converged groups in the shell, then add back the total diffuse mass in the shell as a uniform mass sheet to avoid shot noise effects, and to faithfully represent the total mass in the shell. This procedure requires remapping and bound-checking every particle in the simulation box in every simulation snapshot considered for every light cone produced, which can be very computationally expensive. \\

The fact that no mass outside the light cone field of view is included has two important consequences. First, it has been demonstrated that line of sight material and the large scale density fluctuations of the universe can introduce absolute deflections on the order of arcminutes \citep{uros1994:LSS_lensing}, though these deflections are not directly observable. These large scale modes are not included in our computations due to limits of computational resources and the field of view of the remapped box, which is ultimately a reflection of inadequate simulation volume. As a result, these light cones cannot be perfectly interpreted as the past light cone of some observer. This amounts to the assumption that there are no significant correlations between distant planes, such that in the background close to $z_{\rm max}$ for example, the patch of the universe present in our light cone is statistically consistent with the ``true" patch which would be included had the large scale deflections been computed. Second, a light cone along any random line of sight in the simulation box will appear overdense relative to its (empty) surroundings, leading to an artificially enhanced convergence. To avoid this, the mean matter density of the universe, integrated over the shell volume, must be subtracted from the total mass of each lens plane \citep{birrer2017:los, gilman2019:dm_substructure_fluxratios, meneghetti2020, wagnercarena:2024:substructure, xianzhe2025:los}. This produces a mass \textit{contrast} map for use in the ray tracing. Note that unless the mass map included all mass in the shell (either by using all particles to create the map, or using only converged objects and adding the diffuse mass as a uniform sheet as done here) this subtraction would be an overestimate. Since checking every particle in the box is expensive, approximations such as simply subtracting the mean value of the mass map without explicitly computing the diffuse mass or subtracting the mean matter density (equivalent to assuming the lens plane is exactly consistent with the mean matter density of the universe at its redshift) or using analytical approximations for the total mass in the plane given the mass function of converged groups in that volume, may be desirable. Since we aim to produce a faithful prediction directly from the simulation data, we do not make these approximations and account for all mass explicitly by identifying all the diffuse mass, then adding it to the mass map as a uniform mass sheet. To produce the mass contrast map, we then simply subtract the mean matter density of the universe integrated over the shell from the mass maps, with an equal amount subtracted from each pixel.\\

Unaccounted for thus far is the external shear due to large scale structure outside the light cone. To include this contribution, one could increase the field of view as much as the dimensions of the remapped box would allow, at the expense of significant computational cost. More practical is to sample large scale modes analytically and add a plane of pure shear to each lens plane, for example using GLASS \citep{tessore2023:glass} as implemented including correlations between external convergence and shear and nonlinear corrections in \cite{xianzhe2025:los}. In the light cones produced in this paper we will not include this external shear component, and note that the expected effect is small ($\gamma_{\rm ext}\sim 0.02$, \citealt{xianzhe2025:los}). We describe next in Sec.~\ref{sec:smoothing} how the simulation particle data are smoothed into planes, and detail the procedure by which the stellar light in a chosen filter is estimated in Sec.~\ref{sec:source}. \\

\begin{figure*}
    \centering
    \includegraphics[width=\linewidth]{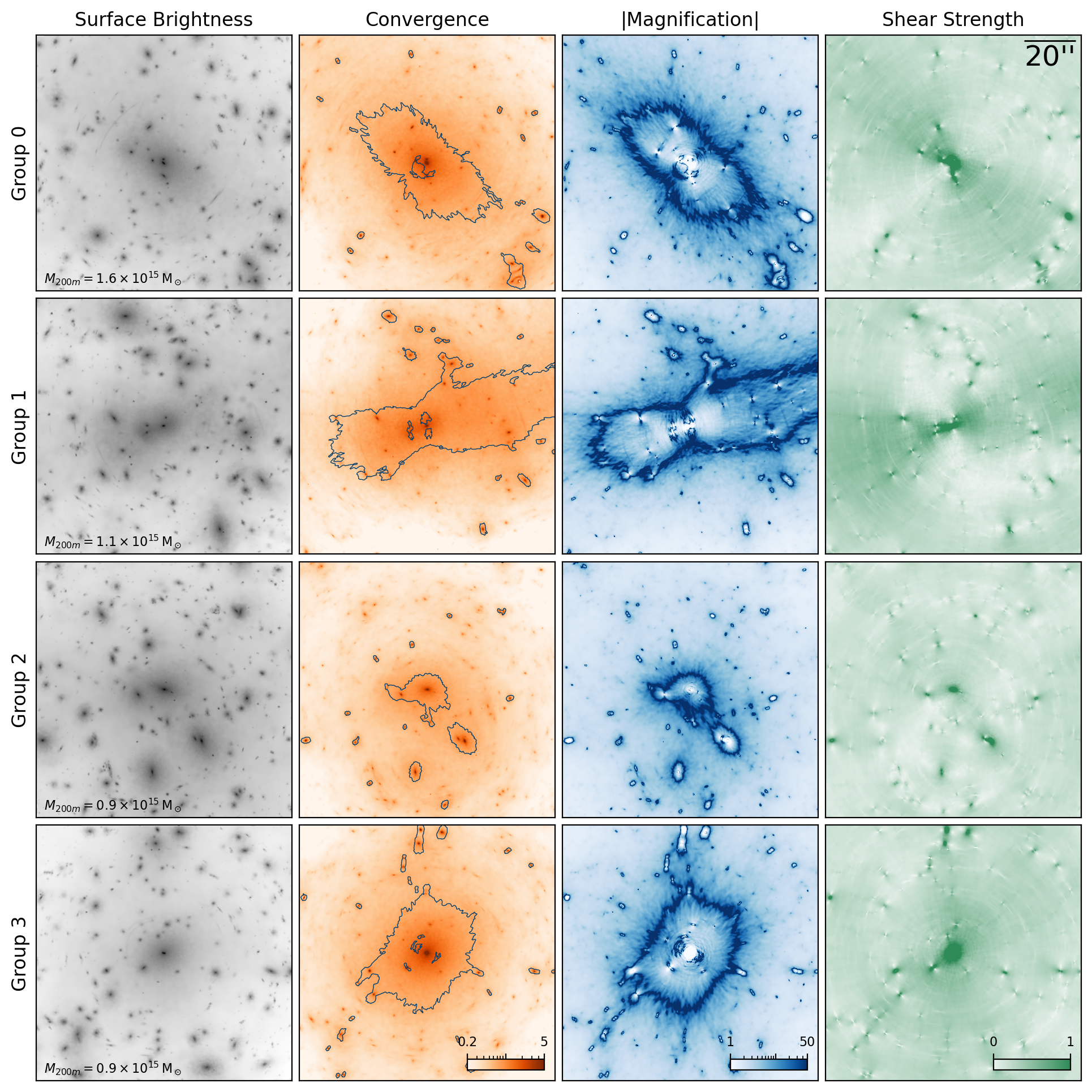}
    \caption{A grid of lensing quantities for light cones utilizing the first 4 groups in the group catalog of TNG300-1 at $z_G=0.38$ as primary lenses in the strong lensing pipeline described in this paper. The columns are (1) surface brightness in ${\rm erg\,s^{-1}\,cm^{-2}\,Hz^{-1}}$ for the whole lensed light cone of each group, computed in the JWST F200W filter but without observational effects, (2) the lensing convergence map $\kappa$ for $z_s=4$, (3) the magnitude of the magnification map $|\mu|$ for $z_s=4$, and (4) the shear strength $\gamma$ for $z_s=4$. In column (1) we annotate the group mass $M_{200m}$ for the primary lens at $z_G$, and in column (2) we show the critical curves for $z_s=4$ in dark blue.}
    \label{fig:lensing_quantity_grid}
\end{figure*}

\subsubsection{Smoothing Procedure}\label{sec:smoothing}
Simulation particles are of finite mass resolution and so must be smoothed to create representative maps of mass or light. This smoothing is typically done adaptively according to the density of particles in an environment, as the more particles in a given area, the more reliably the underlying gravitational potential is sampled. This is typically estimated using a nearest-neighbor calculation to set the scale of a smoothing kernel. Following the procedure of \cite{roche:2024} we create mass maps with the \texttt{SWIFTSIMIO} projection module \citep{borrow2020}, in particular the ``subsampled'' backend, which operates in a similar way to the standard nearest neighbor smoothing but with proportional contribution to pixels and thus more desirable convergence qualities \citep{borrow2021}, especially at low resolutions and specifically in volumes with fewer particles. We use 32 nearest neighbors\footnote{We do not observe a strong dependence of the lensing properties on this choice, but using $\lesssim 20$ neighbors can introduce significant shot noise in the light maps. It is also possible to use more neighbors for the DM smoothing than that for stars to minimize this effect.}, and the default Wendland-C2 kernel \citep{wendland:1995, dehnen:2012} with $\gamma = 1.936$. This smoothing procedure is known to be well-converged and does not artificially dilute the central density of halos, though the central density can be significantly impacted by choices of force softening and mass resolution \citep{borrow2021}.\\

We choose to produce the planes with a constant pixel angular size, such that each plane has the same number of pixels, meaning the comoving sizes (in ckpc) of the pixels in each plane are larger at higher redshifts. We match the angular resolution of the ray tracing grid to this resolution (see Sec.~\ref{sec:raytracing}), and therefore, this choice makes stacking both the lensed and unlensed source planes very simple. We choose to produce images with a resolution of $1024\times1024$, corresponding to a pixel size of $\sim0.1\arcsec$ at the chosen field of view of $100\arcsec$ for the light planes (and thus, $5120\times5120$ for the mass planes). At the chosen cluster redshift of $z_G=0.3802$ this corresponds to a pixel size of $\sim0.5\,{\rm kpc}$, while at $z_{\rm max}$ this corresponds to $\sim 0.7\,{\rm kpc}$. We note that these pixel sizes are both less than the gravitational softening length of the parent simulation ($\epsilon_{z=0}=1.5\,{\rm kpc}$). We notice no significant changes in the images or critical structure when doubling this resolution.

\subsubsection{Source planes}\label{sec:source}
We produce pixellized maps of the flux in each source plane in a manner very similar to the mass planes, but the smoothing weights used for each particle are the appropriately estimated and redshifted fluxes in a chosen observational filter rather than the particle mass. We follow the procedure of \cite{vogelsberger:2020} to compute the emission from the simulation particles and choose not to include dust or nebular emission for simplicity; however, we note that these features could be straightforwardly included by following \cite{vogelsberger:2020}. This procedure models the rest frame SED of a stellar particle using its age, initial mass at formation, and metallicity, then redshifts the SED using the source plane redshift and computes the apparent flux for an observer at $z=0$ in a particular observational band $B$. We accelerate this computation by first evaluating redshifted magnitudes in the $B$ band on a logarithmically-spaced grid in stellar age and metallicity for each snapshot redshift considered. We then evaluate the smoothing weight, i.e., the apparent flux in the desired band for a stellar particle in the source plane at redshift $z_s$ via interpolation of the magnitude grid for that redshift. For details of this procedure and the flux computation, see \cite{vogelsberger:2020}. \\

Since smoothing is necessary to produce faithful representations of the potential sampled by the simulation particles, we cannot include the light from subhalos with less than the number of nearest neighbors used in the smoothing. We use 32 neighbors of a particle for the smoothing here, and therefore we only use the particle data for subhalos with 33 or more star particles when creating the light planes via the smoothing procedure. For the remaining subhalos, we choose to enhance the images using simple analytic replacements. This replacement could be implemented with a higher degree of complexity, using axis ratios and metallicities of the subhalos to produce more varied and complex light distributions. However, for this illustrative example, we use azimuthally symmetric S\'ersic profiles with magnitudes proportional to the number of star particles in the subhalo. We determine these magnitudes by computing the flux per initially formed solar mass of stars in the same manner as in \cite{vogelsberger:2020}, using the median metallicity and age of the star particles in the subhalos with more than 33 star particles in that shell. A more sophisticated treatment may be necessary depending on the intended analysis, such as replacing sources with matched objects extracted from a higher resolution simulation. Here however we choose an effective radius of $R_e=1.5\,{\rm kpc}$ and a S\'ersic index of $n=2$ for all of these profiles. The S\'ersic effective radius is chosen according to the size-mass relation of \cite{CANDELS2014} in their lowest stellar mass bin, for which $R_e$ is approximately constant with redshift for early-type galaxies.  \\

\subsection{Ray Tracing} \label{sec:raytracing}

Once the lens and source planes are created for a particular light cone, one can compute the deflections of light rays coming from each source plane in this cone as they pass through the lens planes at lower redshifts. We compute this lensing result for each source plane separately, then stack the results to produce a combined image (see Fig.~\ref{fig:flowchart}). To perform the ray tracing, we use \texttt{GLAMER} \citep{metcalf2014:glamer,petkova2014:glamer2}, a multi-purpose lensing code written in C++ that can use pixelated source and lens planes as inputs to a grid-based ray tracing algorithm. When using pixelated maps in \texttt{GLAMER}, the lensing quantities (the potential, deflection and shear maps) are computed from the surface density map via a Fast Fourier Transform (FFT) approach. The lensing quantities at arbitrary points on the plane are found by linear interpolation from these maps. We choose to use a nonadaptive, fixed resolution grid that matches the high angular resolution of the planes (Sec.~\ref{sec:planes}).\\

The FFT approach entails some known limitations; for example, it is best to use a grid spacing that is smaller than the particle smoothing scale everywhere, and thus, for a uniform grid, this can lead to long computation times. Another limitation of this approach is that the edges of the grid must be padded with zeros to neutralize the periodic boundary conditions, which further increases the size of the grid. We choose a zero padding factor of 4 times the number of pixels in the unpadded grid. This approach is also unable to include point masses, but since we only include smoothed simulation structures, there are no pointlike objects to represent in this case. Note that an adaptive kernel has been implemented in \texttt{lenstronomy} \citep{birrer2021:lenstronomy} which performs the FFT on two grids with different spacings, significantly reducing the computational demand present in the uniform grid approach. \\

The critical curves, which are the curves in the image plane of formally infinite magnification, and the caustics, which are their counterparts in the source plane, are computed using \texttt{GLAMER}. Since the magnification $\mu$ changes sign across a critical curve, the boundaries of the regions of negative magnification may be identified as the critical curves. In the case of a uniform, nonadaptive grid as used here, \texttt{GLAMER} identifies these boundaries by searching the grid for sign changes in $\mu$. Once a boundary cell is found the algorithm follows the boundary around until it returns to the same cell.  In this way properly ordered and separated closed curves are found \citep{metcalf2014:glamer}. Once a curve is found, its cells are excluded from the pool of boundary cells that are used to construct the next curve.  This is done until there are no more unassigned boundary cells. The areas of these curves are computed by discrete line integral around the segments of the curves, in other words, by adding the signed area of triangles where the vertices are consecutive points on the curve and a fixed central point.  Caustics are found by mapping the critical curves to the source plane. In this paper, the ``primary'' critical curve for a cluster is the single largest area critical curve for that field of view. This definition can be misleading for merging systems with disconnected critical components of approximately equal size, but we do not expect this to have a significant effect on the results discussed here.

\section{Simulation Data}\label{sec:data}
\begin{figure}
    \centering
    \includegraphics[width=\linewidth]{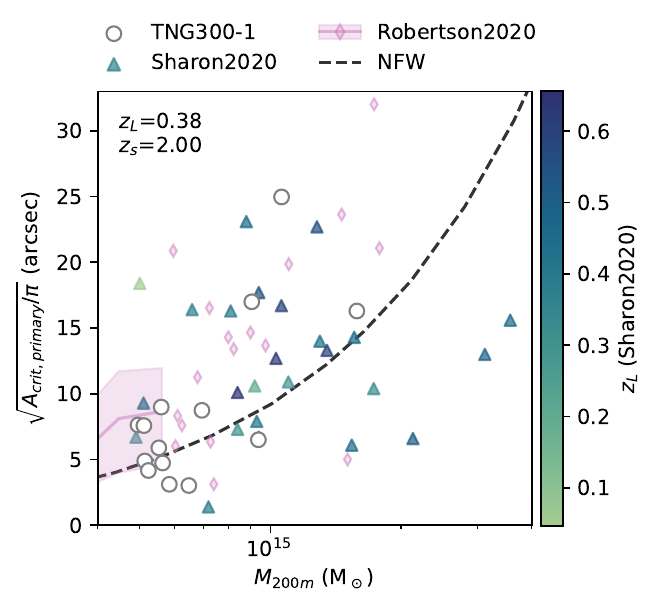}
    \caption{The effective radius of the primary critical curve vs mass for light cones centered on all groups (clusters) above $M_{200m}=5\times 10^{14}\,{\rm M_\odot}$ in the TNG300-1 group catalog at $z_G=0.38$. The source plane redshift  is chosen to be $z_s=2$. We also show the effective radii of the lens models of HST observations presented in \cite{sharon:2020} which describe clusters at various redshifts, each with $z_s=2$. In pink and black respectively are the simulation results of \cite{robertson2020:einstein_radii} and a curve demonstrating the effective radius of an isolated NFW halo with concentration fixed by a mass-concentration relation (see text), both with lens and source plane redshifts matching $z_G$ and $z_s$ for the TNG data.}
    \label{fig:critical_area_vs_mass}
\end{figure}

To utilize the pipeline outlined in Sec.~\ref{sec:methodology}, one must have access to many snapshots of a uniform box cosmological hydrodynamical simulation run to redshift zero, for which stellar ages, initial masses and metallicities are tracked in each snapshot. One must also choose a simulation with a box size sufficiently large to contain galaxy clusters at the desired mass scale, and resolution high enough to resolve background sources, cluster cores, and cluster substructure. A widely-used and publicly available simulation suite satisfying these conditions is IllustrisTNG (TNG; \citealt{springel:2017:ITNG, pillepich:2017:TNGgalaxyformation, marinacci:2018:ITNG, naiman:2018:ITNG, nelson2021:illustristng}). TNG is a suite of cosmological hydrodynamical simulations spanning various simulation box sizes, mass resolutions, and physics models based on the \textsc{AREPO} code \citep{springel:2010:arepo}. This project succeeds the Illustris project \citep{vogelsberger:2014:illustris, vogelsberger:2014:illustrisgalaxies, genel:2014:illustris, sijacki:2015:illustris}, improving upon the galaxy formation model \citep{vogelsberger:2013:illustrismodel} and expanding the scope to larger volumes and higher resolutions. Groups in IllustrisTNG are identified via a friends-of-friends algorithm with a linking length $b=0.2$ in units of the mean interparticle separation \citep{springel:2001} with substructure identified via \textsc{subfind} \citep{dolag:2009:subfind}. The TNG simulation suite consists of TNG50, TNG100 and TNG300, characterized by their varying simulation box sizes of $35$, $75$ and $205\, \rm{cMpc}$$\,h^{-1}$ respectively.\\

We focus on the highest resolution TNG300 box for the present analysis, TNG300-1, as it contains 14 groups above our chosen minimum cluster mass $M_{\rm 200,mean} = 5\times 10^{14}\,{\rm M_\odot}$ at the chosen primary lens redshift $z_G=0.38$. This run employs $2500^3$ dark matter particles and an identical initial number of gas particles, a baryonic mass resolution of  $7.6\e{6}\,\msun\,$$h^{-1}$ and DM mass resolution of $4.0\e{7}\,\msun\,$$h^{-1}$. The gravitational softening length of the stars and DM is $1\,{\rm kpc}\,h^{-1}$ at $z=0$ and the minimum adaptive gas softening length is $0.25\,{\rm ckpc}\,h^{-1}$. The cosmology used in TNG is that of \citealt{planck2016} with parameters $\Omega_{\rm{m}} = \Omega_{\rm{CDM}} + \Omega_{\rm{b}} = 0.3089,\: \Omega_{\rm{b}} = 0.0486,\: \Omega_\Lambda = 0.6911,\: \sigma_8 = 0.8159,\:n_s = 0.9667,\: h = 0.6774$. All computations depending on cosmology, such as the computation of apparent flux of the source planes, are computed consistently with this cosmology.\\

The TNG snapshots are stored in either ``full'' or ``mini'' format, where the mini snapshots contain only a subset of the fields for each particle and are stored more frequently than the full snapshots. The mini snapshots contain the necessary stellar information for the present analysis, and so a total of 80 snapshots at redshifts with $z \leq z_{\rm max}$ are available, where $z_{\rm max} = 4.18$. These snapshots are output at variable spacing in redshift, with a median snapshot spacing of $\Delta z =0.026$ in redshift, minimum spacing of $0.01$ and maximum of $0.163$ for snapshots below $z=3$. This corresponds to a median interplane spacing of $73\,{\rm cMpc}$, a maximum of $223\,{\rm cMpc}$ and a minimum of $43\,{\rm cMpc}$. We also use a minimum plane redshift of $z=0.2$ to generate the example light cones in this paper, as the  separation in redshift between snapshots decreases significantly toward $z=0$, dramatically increasing the computational cost for relatively few additional structures. Note that this means our later estimates of the scatter in critical areas due to line of sight material will be conservative and in fact represent a potential \textit{underestimate} of the magnitude of the effect. The resulting number of planes is $\sim 3$ times the recommended number for convergence in \cite{petkova2014:glamer2}, which suggests using $\sim 24$ planes at $z_{\rm max} = 4.18$.  \\ 

\section{Results}\label{sec:results}
\begin{figure}
    \centering
    \includegraphics[width=\linewidth]{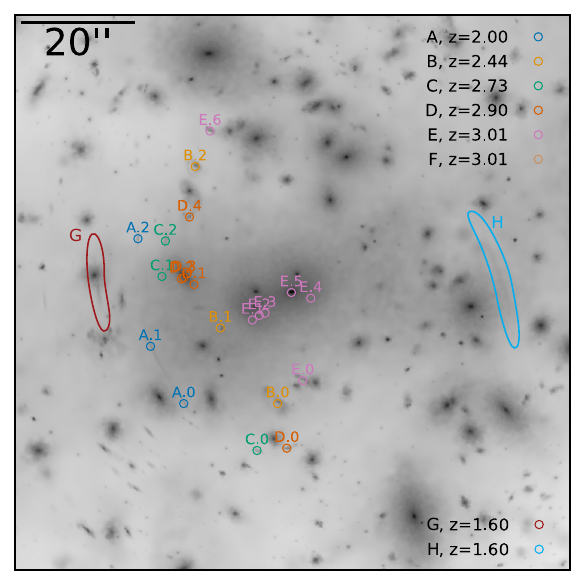}
    \caption{The lensed surface brightness in the JWST F200W band of the light cone corresponding to group 1 in the TNG300-1 group catalog at $z_G=0.38$. Overlaid are all multiply imaged sources in the light cone which contain at least 2 visually identifiable images, along with two large arcs which do not correspond to multiple images. Some multiple images, especially in the cluster core, are not visually distinguishable.}
    \label{fig:multiple_image_example}
\end{figure}

\begin{figure*}
    \centering
    \includegraphics[width=\linewidth]{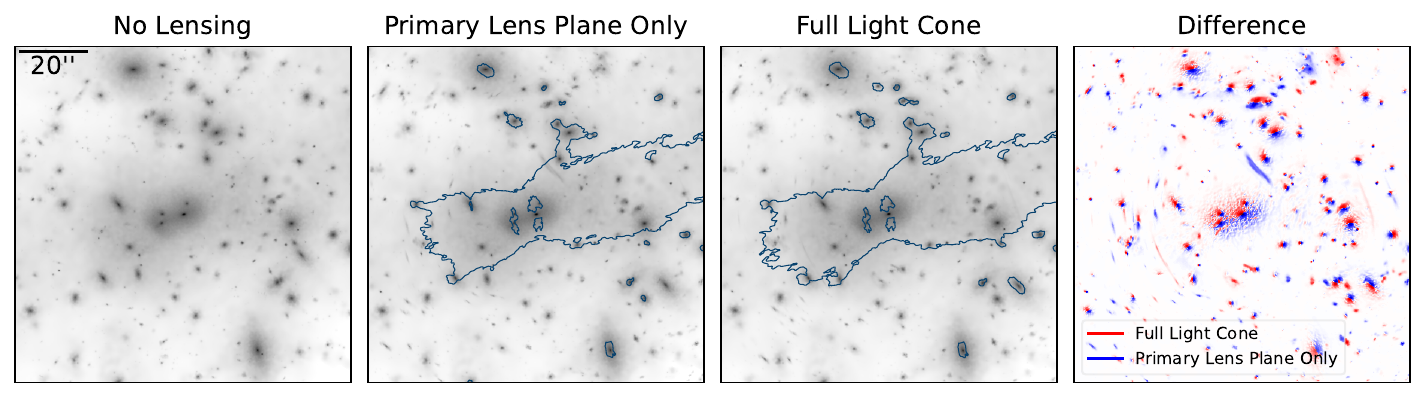}
    \caption{The lensed surface brightness in the JWST F200W band of the light cone corresponding to group 1 in the TNG300-1 group catalog at $z_G=0.38$, shown in the first panel by stacking the unlensed planes (marked as ``No Lensing''). In the second panel, we perform ray tracing of the planes with $z > z_G$ using the mass in the primary lens plane only, labelled ``Primary Lens Plane Only'', and note that this includes both the primary lens cluster and correlated structure within $\sim 35\,{\rm Mpc}$ along the line of sight. We also show the critical curves for $z_s=4$ in dark blue. In the third panel we show the result of lensing each source plane through all available lens planes at lower redshifts in the light cone (the default output of the pipeline described in this paper; labeled as ``Full Light Cone'') and the corresponding critical curves. In the fourth panel, we show the difference between the surface brightness with and without LoS planes, with a diverging symmetric logarithmic colormap diverging from white at 0 difference.}
    \label{fig:LoS_difference}
\end{figure*}

We now generate some example outputs of the pipeline outlined in Sec.~\ref{sec:methodology} using the simulation data of Sec.~\ref{sec:data}. We examine the features of the pipeline outputs in Sec.~\ref{sec:pipeline_outputs}, analyzing the lensing quantities, multiple image characteristics, and critical curves. In Sec.~\ref{sec:LoS}, we examine the effect of line of sight material to study the impact of the common exclusion of the line of sight in lens modeling and sometimes in mock image generation. We determine the magnitude of the deviation of image positions with and without line of sight material in the light cone included, and the corresponding changes to the critical curve structure and supercritical fraction of the sky in the group (cluster) field of view. \\

First, to illustrate how the final lensed images are developed, in Fig.~\ref{fig:individual_planes} we show examples of unlensed light planes forming part of a light cone, and the corresponding lensed planes computed by ray tracing those images through all mass planes at lower redshifts. We also show the final images obtained by summing the source planes at all redshifts, both with and without lensing. The brightness shown here is in the JWST F200W filter. In the top (unlensed) row, we label some sources identified by SourceExtractor \citep{bertin1996:sourceextractor, barbary2016:sep} that correspond to multiple images in the bottom (lensed) row. The cluster is chosen to be at $z_G = 0.38$ in this light cone, and due to a combination of geometric unfavorability and the small amount of material in the foreground of the cluster, the unlensed and lensed panels are nearly indistinguishable. For background planes at more geometrically favorable redshifts, such as $z=2.44$ and $z=3.01$, we see sources being multiply imaged and significant distortion of the surface brightness. In the summed lensed panel, we see various arclets and highly sheared background sources, many from planes other than the example planes shown.\\

\subsection{Lensing Properties}\label{sec:pipeline_outputs}
In Fig.~\ref{fig:lensing_quantity_grid} we show examples of the lensed surface brightness in the JWST F200W filter for light cones centered on the first 4 groups in the TNG300-1 group catalog, with $z_G=0.38$. We also include three lensing quantities, namely the dimensionless mass surface density or ``convergence", the absolute value of the magnification, and shear strength $\gamma = \sqrt{\gamma_1^2 + \gamma_2^2}$ (where $\gamma_{1,2}$ are the two independent components of the shear tensor) for each light cone at $z_s=4$. It is necessary to choose a source plane redshift for visualization of these quantities, unlike the surface brightness panels. We also show the critical curves on the convergence panels, again for $z_s=4$. No instrumental effects have been modeled in the surface brightness column, but mock observations appropriate for comparison to particular observational data could be straightforwardly produced according to instrument characteristics and observation parameters like exposure time, chosen filter, and seeing.\\

Some features of the images that are readily apparent are as follows. First, the critical curve for group 1 is significantly elongated, which is due to an ongoing merger. Second, there are some cloudy and unconverged-appearing features in the surface brightness panels, for example, in the center right of the group 1 light cone, corresponding to structures with a small number of star particles and thus suffering from shot noise effects, or structures with a large smoothing length. The number of poorly converged objects with visible shot noise effects is significantly reduced by our cuts on the total and stellar mass of included subhalos, and the number of overly-smoothed structures is reduced by changing the number of nearest neighbors used in the smoothing calculation. However, if this number is too small, the shot noise effects in low-density regions such as the outskirts of halos become more apparent. Using analytic replacements for those subhalos with less star particles than the number of neighbors for smoothing mitigates this issue, but a more consistent approach may involve replacement with higher-resolution simulation data or adopting zoom techniques. \\

Second, we observe several large tangential arcs in the group 0 and 1 light cones, consistent with expectations from CLASH clusters for example, which have been observed to contain $\sim 4$ arcs over $6\arcsec$ in length per cluster \citep{xu2016:giant_arc_statistics}. Group 3 exhibits some long but faint arc structure, and group 2 contains no obvious lensing evidence, though it is noteworthy that group 2 has a noticeably different morphology than the other examples, with two smaller primary critical curves. Since no observational effects have been modeled at this juncture, in principle, one may expect an equal or greater number of giant arcs in these simulated images than that in deep observational samples. The rate of giant arcs is a product of many factors, including the critical area of the cluster, observational parameters, selection functions, the source object number density, the luminosity function of sources, and the surface brightness profiles of those sources \citep{meneghetti2000:cluster_arcs, meneghetti2003:bcg_lensing, meneghetti2007:arc_sensitivity, meneghetti2007:cluster_lensing_density_profiles, meneghetti2010:marenostrum_lensing_biases, torri2004:merger_arc_statistics}. It has been well established that the galaxy stellar mass and luminosity functions of TNG closely match observations at redshifts lower than our $z_{\rm max}$ of $\sim 4$ \citep{pillepich2018:tng_stellar_mass, vogelsberger:2020}, and while a detailed study of the consistency of observed and simulated lensing statistics is outside the scope of this paper, as an initial consistency check we examine the cluster critical areas in Fig.~\ref{fig:critical_area_vs_mass}. Here we show the effective radius of the primary critical curve area as a function of the mass of the primary lens for both our data and observations and simulations in the literature. We define the primary critical curve as the curve with the largest area, which is representative in most cases but can be problematic for primary lenses like Group 2. We use a cluster lens redshift of $z_G=0.38$ and source plane redshift of $z_s=2$ to remain consistent with the other datasets where possible in this Figure, and examine light cones centered on all groups in the TNG300-1 group catalog at $z_G$ with mass $M_{200m} \geq 5\times 10^{14}\,{\rm M_\odot}$. Here, we see that the primary critical curve area is broadly consistent with both existing observations from lensing-selected clusters \citep{sharon:2020}, simulations not including a line of sight \citep{robertson2020:einstein_radii}, and an isolated NFW halo of variable mass and concentration set according to the mass--concentration relation of \cite{ludlow2016:mass_concentration}, similar to the presentation of \cite{robertson2020:einstein_radii}. This consistency is in large part due to the large scatter of primary critical curve area with cluster mass, as a result of important effects such as cluster orientation and ellipticity as well as the scatter in the mass-concentration relation. Without line of sight material but including selection effects and simplified analytical sources, \cite{robertson2020:einstein_radii} report that their data admit a giant arc rate consistent with observations \citep{wiesner2012:sdss_strong_lensing}. To appropriately determine the giant arc rate in our data with simulation sources, one must model observational effects and account for the truncation of the mass function due to simulation mass resolution in both the lenses and the sources of the light cone. To then determine if that rate is in tension with observations, one must utilize an observational dataset with minimized selection effects. Such a full analysis is outside the scope of this paper, as here we focus on presenting this new methodology, but it is promising that the \cite{robertson2020:einstein_radii} effective Einstein radii are consistent with our data.\\

We also observe various multiply imaged source systems in these images, some of which are identifiable by eye upon close inspection and are often found more easily by stacking images in blue and red filters. To systematically mark the multiply imaged sources, we first use SourceExtractor \citep{bertin1996:sourceextractor, barbary2016:sep} on every source plane in a light cone to identify every source object's position. We then ray trace all these positions and identify the sources that are multiply imaged, tagging those multiple images. In Fig.~\ref{fig:multiple_image_example} we show the image locations of all multiply imaged sources in the light cone corresponding to group 1 in the TNG300-1 group catalog at $z_G=0.38$ out to $z_s=4$ which have at least 2 visually-identifiable images. Note again that we do not model observational effects here. We label the images corresponding to one source object with a letter, and the multiple images in that system with a number, beginning with zero. One can see configurations with between 3 and 5 images, and some large arcs labeled G and H which are not multiply imaged. Catalogs of multiply imaged systems are straightforward to produce in these light cones, and could be used to study the systematics of image identification (e.g., \citealt{johnson2016:sl_systematics}), biases in lens modeling inferences (e.g., \citealt{gonzalez2012:sl_core_mass}), and potentially contribute to future data challenges produced using simulation data in a more unsupervised manner than previous iterations \citep{meneghetti2017:challenge}. \\

\subsection{Impact of the Line of Sight}\label{sec:LoS}
We now examine the difference in the lensing features of our light cones when including line of sight structure, compared to the common assumption of a single lens plane at the cluster redshift. Note that the thickness of the primary lens plane here is determined by the simulation snapshot output frequency, and corresponds to a depth $\Delta z \simeq 0.02$ ($\sim 70\, {\rm Mpc}$), which is much larger than the region occupied by the particles identified as the cluster group in the simulation ($\sim$ a few Mpc), but smaller than the depths often used to define observed clusters, which are typically around $\Delta z \simeq 0.05$ ($\sim 200\, {\rm Mpc}$; \citealt{niemiec2023:weakstrong}) but can be as large as $\Delta z \simeq 0.2$ ($\sim 600\, {\rm Mpc}$; \citealt{gledhill2024:canucs_mass_model}). As a result, correlated structure is already included in the primary lens plane, and we choose to study the effect of uncorrelated structure here. In this section, ``with LoS'' or ``full light cone'' refers to performing the ray tracing using all mass planes, and ``no LoS'' or ``primary lens plane only'' refers to performing the ray tracing using the primary lens plane only, i.e. including correlated but not uncorrelated matter in the ray tracing. \\

In Sec. \ref{sec:image_positions} we examine the difference in lensed source positions with and without the inclusion of uncorrelated line of sight structure. We then focus on changes to the critical curves and total critical area on the sky due to line of sight material in Sec. \ref{sec:crit_structure}.

\subsubsection{Image Positions}\label{sec:image_positions}
In Fig.~\ref{fig:LoS_difference} we show the summed surface brightness of the light cone corresponding to group 1 in the TNG300-1 group catalog at $z_G=0.38$ (as in Fig.~\ref{fig:lensing_quantity_grid} and \ref{fig:multiple_image_example}) but with 3 different distributions of mass in the light cone. First, we show the unlensed stacked result, then the result of stacking the source planes lensed through only the mass of the plane at the cluster redshift, and finally the result when each plane is lensed through the mass of every plane at lower redshifts (the default output of the pipeline described in this paper). \\

We also show the critical curves, from which it is clear that the uncorrelated line of sight structure slightly increases the size of many critical curves in this example, and introduces additional supercritical structure as will be discussed further in the next section. In the final panel, we show the difference between the full light cone and primary lens plane only cases. This difference panel highlights the change in position and shape of various images and arcs, or the existence of certain images in the full light cone map that do not appear in the no LoS case. A small shift ($\sim 2 \arcsec$) in the position of many cluster member galaxies is present due to deflection by foreground mass. Some lensed images in this example change position by almost $\sim 10 \arcsec$, such as the arcs in the center-left and center-right portions of the image. This difference is significantly larger than typical strong lens mass modeling image residuals \citep[$\lesssim 1\arcsec$;][]{mahler2018:sl_model, bergamini2019, bergamini2021, bergamini2023, bergamini2023:MACS, sharon:2020, cerny2026:slice}, but the actual impact on the quality of lens modeling inference is likely to be negligible, and will be examined in future work. Some images change in their \textit{relative} position by $\sim 2-4$ arcseconds (comparing the separation between two red images to that between the corresponding two blue images), such as those in the center left of the image. This aligns with the expected relative deviations between images due to line of sight matter \citep{host2012:los_lensing, raney2020:los_lensing}. This example highlights the fact that the uncorrelated line of sight can have a significant effect on cluster lensing, and performing lens modeling assuming that the effect of uncorrelated structure can be absorbed into the cluster plane will result in systematic biases. A detailed study of the degree to which this can affect lens models and the derived integrated lensing masses will be examined in a follow-up paper, and we treat this primarily as an illustrative example. \\

\begin{figure}
    \centering
    \includegraphics[width=\linewidth]{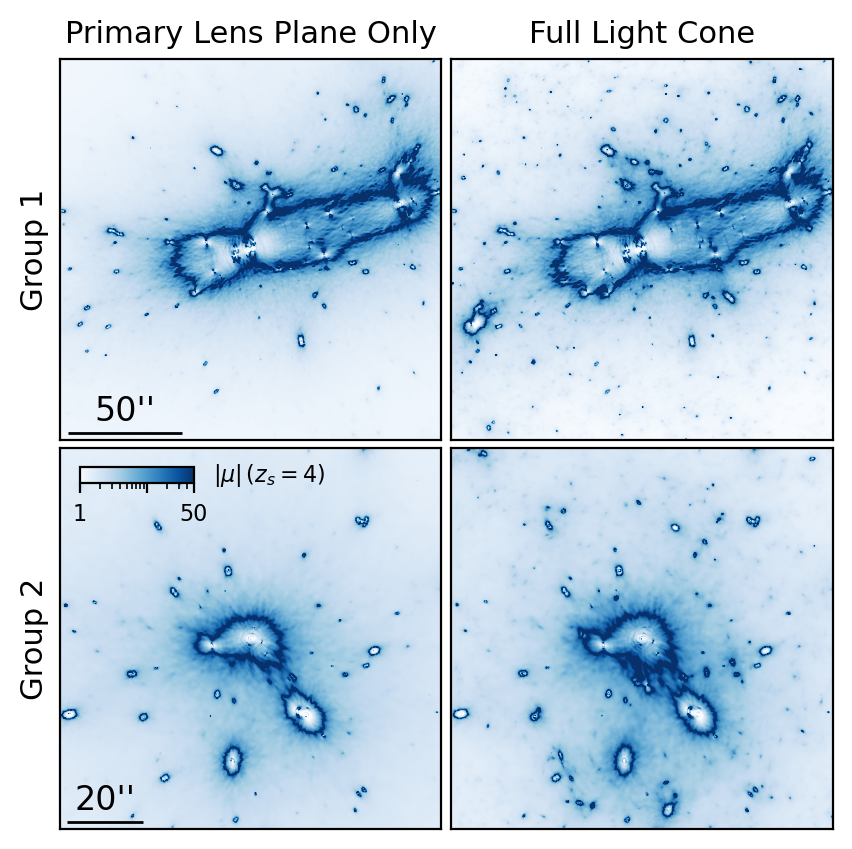}
    \caption{(\emph{Left Column}) The absolute magnitude of the magnification due to groups 1 and 2 in the TNG300-1 group catalog at $z_G=0.38$, including correlated structure within $\sim 35\,{\rm Mpc}$ along the line of sight. (\emph{Right Column}) The absolute magnitude of the magnification in the light cone corresponding to groups 1 and 2 in the TNG300-1 group catalog at $z_G=0.38$, including all line of sight material. In all panels we choose $z_s=4$ for the source plane, and note that the field of view is larger in the top row.}
    \label{fig:LoS_magnification}
\end{figure}

\subsubsection{Critical Curve Structure}\label{sec:crit_structure}
We now examine the changes in critical curve structure when line of sight material is included in the lensing calculations. Already in Fig.~\ref{fig:LoS_difference} one can see that a number of line of sight galaxies close to the cluster redshift, which could be identified as cluster members in observations with wider definitions of cluster depth, make a significant difference to the primary critical curve shape and number of secondary curves. The difference in the density and sizes of critical curves in the LoS and no LoS cases is made clearer by looking at the magnification maps, of which we show two examples in Fig.~\ref{fig:LoS_magnification}. Here it can be seen that the highly magnified region (dark blue, corresponding to $|\mu|\geq 50$) changes shape in particular to the left for group 1 and the bottom for group 2, and the number density of secondary lensing features is significantly enhanced.\\

Some critical curves present in the full light cone have no counterpart in the single lens plane case, and correspond to line of sight objects. The curves already present in the single plane problem are however enhanced in size and modified in shape by inclusion of the line of sight material. We make this difference more explicit in  Appendix Fig.~\ref{fig:crit_radius_grid}, in which we show the distribution of effective radii of critical curves when line of sight material is included or excluded. We show these distributions for light cones centered on the groups in the TNG300-1 group catalog at $z_G=0.38$ with $M_{200m} \geq 5\times 10^{14}\,{\rm M_\odot}$, and a source plane redshift of $z_s=4$, where $M_{200m}$ is the mass contained within a sphere with average density equal to 200 times the mean matter density of the universe at that redshift. What is readily apparent in these panels is that the line of sight material adds significant lensing power to the cluster, in particular contributing a large number of lensing features at the galaxy scale and below ($\sqrt{A_{\rm crit}/\pi}\lesssim 1''$), though this does depend on the choice of field of view. This insight invites the study of how line of sight material impacts an outstanding tension in the field of simulated cluster strong lensing, namely the galaxy-galaxy strong lensing (GGSL) cross section \citep{meneghetti2020, meneghetti2022, meneghetti2023, tokayer2024, dutra2025:ggsl}. It is to be noted however that when computing GGSL cross sections, it is standard to include only those critical curves associated with identified cluster members, and so the effect of the line of sight would primarily be in the enhancement of the sizes of those curves, not necessarily their number. This again depends on the depth in redshift used to define the ``cluster''. We will examine the impacts of the line of sight on the GGSL cross section in a future work. \\

\begin{figure}
    \centering
    \includegraphics[width=1.0\linewidth]{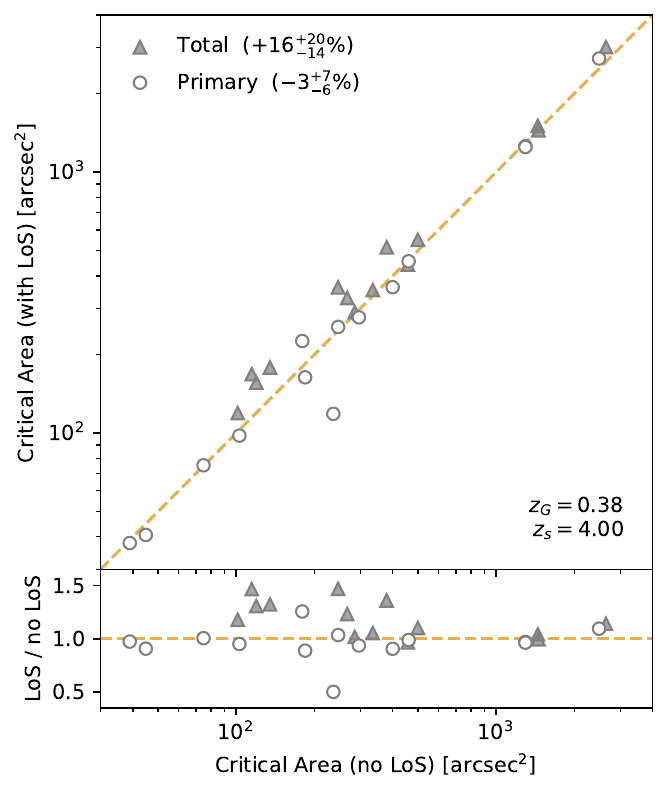}
    \caption{Comparison of the primary critical curve area and total critical area on the sky within $100\arcsec$ of the cluster potential minimum both with and without line of sight material. We use light cones centered on all groups with $M_{200m}\geq5\times 10^{14}\,{\rm M_\odot}$ in the TNG300-1 group catalog at $z_G=0.38$. In these light cones, the target cluster is chosen to be at redshift $z_G=0.38$ and the critical curves are computed for a source plane redshift of $z_s=4$. Shown in both panels is a dashed line corresponding to equal critical areas with and without the line of sight material. ``No LoS'' means ray tracing is performed using only the primary lens plane (cluster $\pm \sim 35\,{\rm Mpc}$), and ``with LoS'' means ray tracing is performed with all planes.} 
    \label{fig:pcrit_area_ratio}
\end{figure}

What is not clear in Fig.~\ref{fig:crit_radius_grid} is the factor by which critical areas are enhanced. We examine this in Fig.~\ref{fig:pcrit_area_ratio} by showing the total critical area of each light cone within 100$\arcsec$ of the cluster potential minimum both with and without the line of sight material. We do not show enhancements of individual critical curve sizes because there is not an injective mapping from critical curves in the full light cone case to the single lens plane case, since some curves have no counterpart and other curves merge or separate. Here we see that the line of sight material typically enhances the total critical area, with a median and 16/84th percentiles of $16^{+20\%}_{-14\%}$, but in some cases in fact decreases the total critical area.  Some of this boost arises from the addition of new curves due to individual line of sight objects, and some of the enhancement arises from the increase in size of critical curves already present in the single plane lensing case due to the contribution of the additional enclosed mass along the line of sight. There also appears to be a greater fractional effect for smaller objects, which is expected since the amount of added critical structure from the uncorrelated line of sight should not depend on the properties of the cluster. One object sees a $\sim 50\%$ decrease in primary critical area, which we verify is the result of a merging system whose two components are connected by a thin ($\sim 0.3\arcsec$) bridge, which in the full light cone case happens to break, approximately halving the area of the single largest critical curve. We also see in Fig.~\ref{fig:pcrit_area_ratio} that the uncorrelated line of sight objects introduce a $\sim 6\%$ scatter in the area of the primary critical curve in this sample, with one curve enhanced in area by 30\% and another decreased by 50\%. This indicates that these typically unmodeled objects could have a significant effect on the inferred features of the cluster primary lens, since for a point-like lens the area of the critical curve scales linearly with mass.\\

\section{Conclusions}\label{sec:conclusions}

In this paper we demonstrate that the necessary tools are now available to produce strong lensing predictions from cosmological simulations, for which all mass and light information are drawn consistently from a single cosmological volume. We apply this method to the largest groups in the widely-used and publicly-available IllustrisTNG300 box, creating light cones out to $z_{\rm max}\simeq 4$. We demonstrate that these light cones produce typical strong lensing features such as multiple image configurations and arcs without the injection of sources, but do not explicitly determine the detectability of those features or model observational effects in this work. To our knowledge, this is the first demonstration of cluster strong lensing images generated self-consistently from a cosmological hydrodynamical simulation light cone.\\

This methodology does exhibit some limitations, such as the fact that it does not introduce large scale power in the matter power spectrum which is not present in the initial simulation box, meaning the resulting image statistics represent a biased sampling of the large scale power spectrum. This problem could be addressed in various ways by intervening in the creation of the light cone, for example sampling the large scale power spectrum similar to the approach often taken in the literature \citep{wagnercarena:2024:substructure}, but using simulation data to construct the resulting planes rather than analytic profiles. It is also ultimately limited by simulation resolution, and the challenge of requiring very large mass scales for the primary lens ($\sim 10^{15}\,\msun$), and very small mass scales for the substructure and sources ($\sim 10^{10}\,\msun$). This problem could be addressed in future studies by using zoom-in techniques, or carefully mixing data from simulations run with different box sizes and mass resolutions. Another limitation resulting from the finite resolution is that the diffuse mass must be treated differently than converged objects to avoid shot noise effects. We have chosen to approximate the diffuse mass as a uniform sheet, but other approaches such as aggressively smoothing the diffuse particles into a separate mass map which is added to the lens plane would preserve some additional structure. Additionally, to generate a single light cone in which ray tracing can be performed to produce one stacked lensing image, this pipeline requires one to perform computations with the positions of every particle in the simulation box at every available snapshot (up to $z_{\rm max}$), which can be computationally expensive for very large modern cosmological simulations, especially if many light cones are desired. This task is trivially parallelized, but it is noteworthy that with a constant grid size of 5120$\times$5120, saving the required lens, source and lensing quantity planes for each redshift in a single light cone amounts to many gigabytes of files, scaling as $\mathcal{O}(n^2)$ with image resolution, such that for a large catalogue of images very significant disk space and in particular RAM for ray tracing is required. Finally, if generating a large catalog of images from a single simulation box, some structures may be present in more than one image, with a repetition probability dependent on the chosen remapped box side lengths and field of view. However choosing a remapping and field of view which produces light cones that do not take up a significant fraction of the volume of the remapped box, as is appropriate for strong lensing, minimizes this probability. \\

This method of producing image predictions consistently from simulation data shows significant promise for the interpretation of observed strong lens systems, and in understanding the true theoretical predictions of the standard cosmological model for lensing statistics and features. 
Applying lens modeling techniques to mock data has been incredibly successful in understanding systematic biases of lens modeling choices \citep{meneghetti2017:challenge, remolina2020, remolina2021}, and this method represents a more automated and hands-off means of producing large catalogs of such images across a wide range of astrophysical contexts, and from any appropriate cosmological hydrodynamical simulation that one chooses. \\

We conservatively estimate that the uncorrelated line of sight material can have a significant effect on the critical structure, introducing a $\sim 6$\% scatter in the area of the primary critical curve, with an effect as large as $50\%$ even in this small sample of 14 light cones. We find that uncorrelated structure can shift the relative positions of lensed images by several arcseconds, in line with the expectation value of the relative deflection computed from the matter power spectrum \citep{host2012:los_lensing}. These systematics could introduce biases in the inference of primary lens properties such as cluster mass, or in cosmological inference utilizing cluster time delay cosmography, for example. We next intend to use this apparatus to study these questions, in addition to the galaxy-galaxy strong lensing cross section discrepancy in clusters, and biases in understanding dark matter in cluster cores by studying strong lensing/galaxy offsets in simulations and observations. \\

\section*{Acknowledgments}
The authors thank Kiyoshi Masui, Meredith Neyer, and Ethan Nadler for valuable discussions. The authors also thank the anonymous reviewers for their helpful comments.

This work makes use of the following software: \textsc{Python} \citep{python}, 
\textsc{numpy} \citep{numpy:2020}, \textsc{scipy} \citep{scipy:2020}, 
\textsc{astropy} \citep{astropy:2013, astropy:2018}, 
\textsc{jupyter} \citep{jupyter}.

\bibliography{bibliography}{}
\bibliographystyle{aasjournal}

\begin{appendix}
\section{Critical Structure}\label{app:crit}
\begin{figure*}
    \centering
    \includegraphics[width=\textwidth]{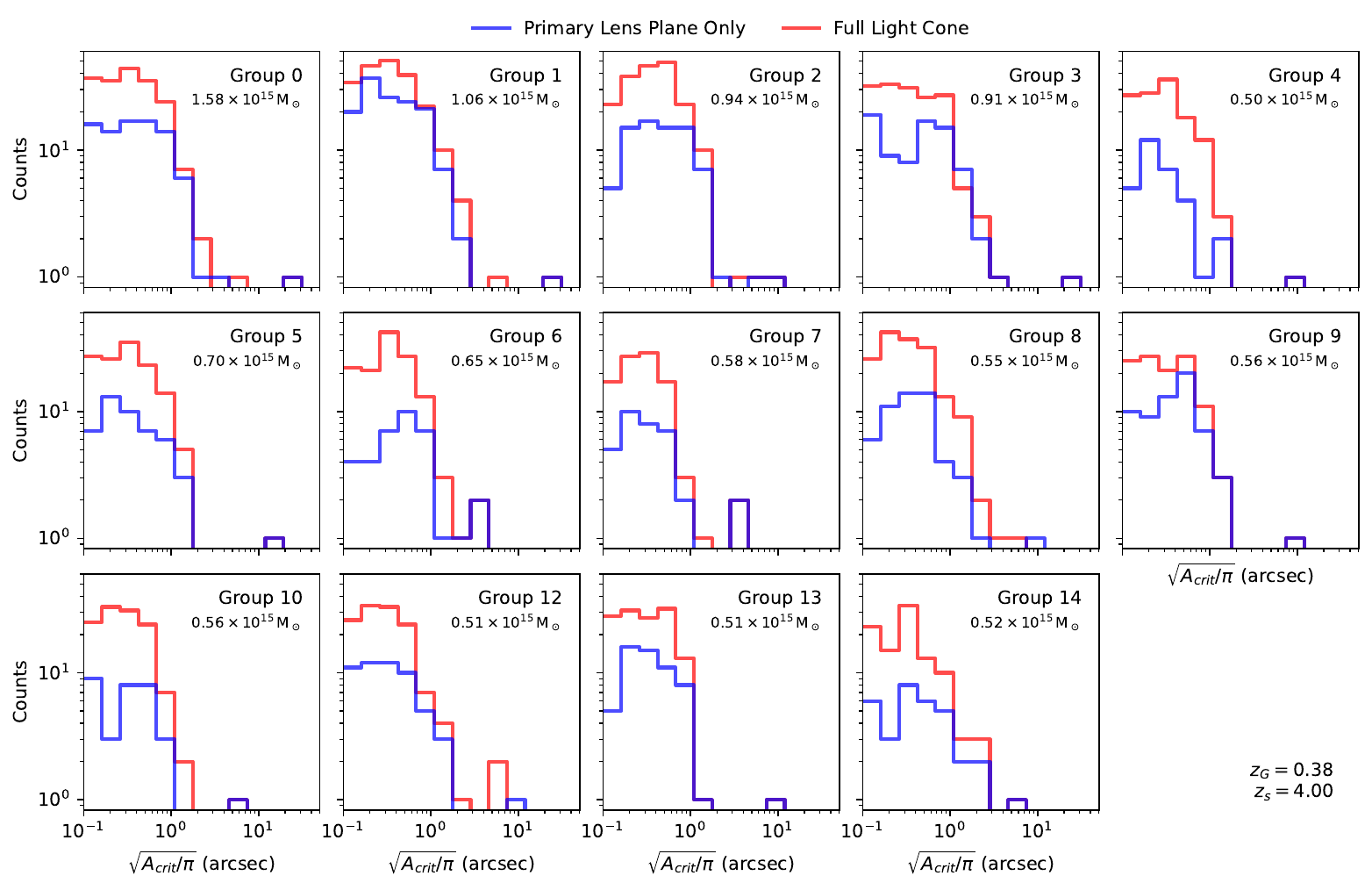}
    \caption{The distribution of critical curve areas within $100\arcsec$ of the cluster potential minimum for light cones centered on all groups (clusters) with $M_{200m}\geq5\times 10^{14}\,{\rm M_\odot}$ in the TNG300-1 group catalog at $z_G=0.38$, both with and without line of sight material. In these light cones, the target cluster is chosen to be at redshift $z_G=0.38$ and the critical curves are computed for a source plane redshift of $z_s=4.0$. The annotated mass in each case is $M_{200m}$ for the primary lens group at $z_G$.}
    \label{fig:crit_radius_grid}
\end{figure*}

Here we show the changes in critical curve structure when including line of sight material in the full light cone, as opposed to the single-plane lensing problem. This is shown in Fig.~\ref{fig:crit_radius_grid}, and discussed primarily in Sec.~\ref{sec:crit_structure}. 
\end{appendix}

\end{document}